# Comment on "Cartesian expressions for surface and regular solid spherical harmonics using binomial coefficients and its use in the evaluation of multicenter integrals"

I.I. Guseinov

*Department of Physics, Faculty of Arts and Sciences, Onsekiz Mart University, Çanakkale,*

**Abstract**

Recently published formulas for the surface and regular solid spherical harmonics and for the expansion of the product of two normalized associated Legendre functions with different centers in ellipsoidal coordinates (Telhat Özdoğan, Metin Orbay, Czech.J.Phys., 52(2002)1297) are critically analyzed. It is demonstrated that the presented in this work formulas are not original and they are available in the literature or can easily be obtained from the published in the literature formulas by changing the summation indices.

Özdoğan and Orbay in Ref.[1] published the cartesian expressions for surface and regular spherical harmonics and the expansion formulas for the product of two normalized associated Legendre functions both with different centers in ellipsoidal coordinates. Eqs.(14),(15), (16), (21), (22), (23) and (27), (28) presented in Ref.[1] by these authors have been already given in Refs.[2] and [3], respectively:

Eqs.(5.4.4), (5.4.5) and (5.4.6) in Ref.[2]

$$Y_{lm}(\pi-\theta,\varphi) = (-1)^{l+m} Y_{lm}(\theta,\varphi) \tag{1}$$

$$Y_{lm}(\theta,\pi+\varphi) = (-1)^{m} Y_{lm}(\theta,\varphi) \tag{2}$$

$$Y_{lm}(\pi-\theta,\varphi+\pi) = (-1)^{l} Y_{lm}(\theta,\varphi), \tag{3}$$

Eqs.(1) and (21) in Ref.[3]

$$F_m(n) = F_m(n-1) + F_{m-1}(n-1) \tag{4}$$

$$F_m(N,N') = \sum_k (-1)^k F_{m-k}(N) F_m(N'). \tag{5}$$

It is easy to show that Eqs.(7),(11),(13),(17) and (18) of Ref.[1] for cartesian expressions os SSH and RSSH can be derived from the published in Ref.[2] formulas by changing the summation indices:

Eqs.(5.2.20) and (5.1.16) in Ref.[2]

$$Y_{lm}(\theta,\varphi) = e^{im\varphi}\left[\frac{2l+1}{4\pi}(l+m)!(l-m)!\right]^{1/2}(\cos\theta)^l \sum_{s=|m|,|m|+2,\ldots}(-1)^{\frac{s+m}{2}}\frac{1}{(s+m)!!(s-m)!!}\frac{(tg\theta)^s}{(l-s)!} \quad (6)$$

$$r^l Y_{lm}(\theta,\varphi) = \left[\frac{2l+1}{4\pi}(l+m)!(l-m)!\right]^{1/2}\sum_{pqr}\frac{1}{p!q!r!}\left(-\frac{x+iy}{2}\right)^p\left(\frac{x-iy}{2}\right)^q z^r, \quad (7)$$

where $p+q+r=l$, $p-q=m$.

The published in Ref.[1] Eqs.[24] and (25) for the expansion of the product of two normalized associated Legendre functions in ellipsoidal coordinates and their use in the evaluation of multicenter integrals over STOs were critically analyzed in our Comment [4]. We have proved that Eqs.(24) and (25) of Ref.[1] are obtained from the expansion relationships contained in our articles [3,5,6] by changing the summation indices.

Thus, the formulas published by Özdoğan and Orbay for cartesian expressions for surface and regular solid spherical harmonics and for expansion of the product of normalized associated Legendre functions in ellipsoidal coordinates are not original and they are available in the literature.